\documentclass[fleqn,10pt]{wlscirep}
\title{Evidencing Unauthorized Training Data from AI Generated Content using Information Isotopes}

\author[1,2$\dag$]{Tao Qi}
\author[2$\dag$]{Jinhua Yin}
\author[3]{Dongqi Cai}
\author[4]{Yueqi Xie}
\author[2]{Huili Wang}
\author[2]{Zhiyang Hu}
\author[2]{Peiru Yang}
\author[1]{Guoshun Nan}
\author[5]{Zhili Zhou}
\author[1*]{Shangguang Wang}
\author[6]{Lingjuan Lyu}
\author[2*]{Yongfeng Huang}
\author[3*,7]{Nicholas D. Lane}
\affil[1]{School of Computer Science, Beijing University of Posts and Telecommunications}
\affil[2]{Department of Electronic Engineering, Tsinghua University}
\affil[3]{Department of Computer Science and Technology, University of Cambridge}
\affil[4]{Department of Computer Science and Engineering, Hong Kong University of Science and Technology}
\affil[5]{School of Artificial Intelligence, Guangzhou University}
\affil[6]{Sony AI}
\affil[7]{Flower Labs}
\affil[*]{Correspondence: sgwang@bupt.edu.cn, yfhuang@tsinghua.edu.cn,
ndl32@cam.ac.uk
}
\affil[$\dag$]{Equal Contribution}
\affil[Remark: ]{This paper presents a preliminary version of our research. The full research content, code and data will be released.}

\usepackage{CJKutf8}
\usepackage{bm}
\usepackage{hyperref}
\usepackage{lineno}
\usepackage{verbatim}
\usepackage{booktabs} 
\usepackage{algorithm}
\usepackage{epsfig}
\usepackage{color}
\usepackage{amsmath}
\usepackage{graphicx,subfigure}
\usepackage{multirow}
\usepackage{makecell}
\usepackage{amssymb}

\usepackage{xr}
\usepackage[mathscr]{eucal}
\usepackage{url}
\usepackage{algpseudocode}

\newtheorem{lemma}{Lemma}

\usepackage{subfigure}

\begin{abstract}

In light of scaling laws, many AI institutions are intensifying efforts to construct advanced AIs on extensive collections of high-quality human data. 
However, in a rush to stay competitive, some institutions may inadvertently or even deliberately include unauthorized data (like privacy- or intellectual property-sensitive content) for AI training, which infringes on the rights of data owners.
Compounding this issue, these advanced AI services are typically built on opaque cloud platforms, which restricts access to internal information during AI training and inference, leaving only the generated outputs available for forensics.
Thus, despite the introduction of legal frameworks by various countries to safeguard data rights, uncovering evidence of data misuse in modern opaque AI applications remains a significant challenge.
In this paper, inspired by the ability of isotopes to trace elements within chemical reactions, we introduce the concept of information isotopes and elucidate their properties in tracing training data within opaque AI systems. 
Furthermore, we propose an information isotope tracing method designed to identify and provide evidence of unauthorized data usage by detecting the presence of target information isotopes in AI generations.
We conduct experiments on ten AI models (including GPT-4o, Claude-3.5, and DeepSeek) and four benchmark datasets in critical domains (medical data, copyrighted books, and news). 
Results show that our method can distinguish training datasets from non-training datasets with 99\% accuracy and significant evidence (p-value$<0.001$) by examining a data entry equivalent in length to a research paper.
The findings show the potential of our work as an inclusive tool for empowering individuals, including those without expertise in AI, to safeguard their data rights in the rapidly evolving era of AI advancements and applications.

\end{abstract}
\begin{document}
\begin{CJK*}{UTF8}{gbsn}
% \linenumbers
\flushbottom
\maketitle

\section*{Introduction}

Artificial intelligence (AI) technologies, exemplified by large language models (LLMs), have witnessed remarkable advancements, showcasing exceptional capabilities in generating content with human-like coherence~\cite{buschoff2024visual,seamless2025joint,dathathri2024scalable}. 
Leveraging these advanced techniques, prominent AI platforms such as ChatGPT and Claude have emerged as effective tools for assisting users in a diverse range of tasks, like daily information seeking~\cite{singhal2025toward,heimberg2024cell,hollmann2025accurate}, auxiliary medical diagnosis~\cite{liu2025generalist,griot2025large,xiang2025vision,zhang2024generalist,zhou2024pre}, content creation~\cite{tanno2024collaboration,bluethgen2024vision,theodorou2023synthesize,van2024adapted}, and scientific discovery~\cite{riveland2024natural,tikochinski2025incremental,kang2024comprehensive,benegas2025dna,deng2024nanobody,shen2024accurate}.
These AI systems have demonstrated the ability to provide expert-level insights tailored to user needs, thereby exerting a growing influence across various societal domains~\cite{liu2025quantitative,wu2024leveraging,fradkin2024latent,luo2024large,lu2024visual,chenthamarakshan2023accelerating}.
Recent investigations into the underlying mechanisms of AI systems have revealed that their primary source of intelligence lies in the extensive knowledge embedded within their training data~\cite{elmopaper,jiao2020tinybert,du2022glm,touvron2023llama}.
Building upon the cumulative achievements in AI development, the principle of scaling has been proposed to underscore the benefits of increasing both model size and dataset volume for enhancing AI performance~\cite{kaplan2020scaling}. 
This scaling law has become a foundational principle in modern AI development, guiding the creation of advanced systems capable of demonstrating emergent super intelligence~\cite{brown2020language,weifinetuned,touvron2023llama,fu2025foundation,wang2024self}.

\begin{figure}[!h]
    \centering
    \includegraphics[width=0.96\linewidth]{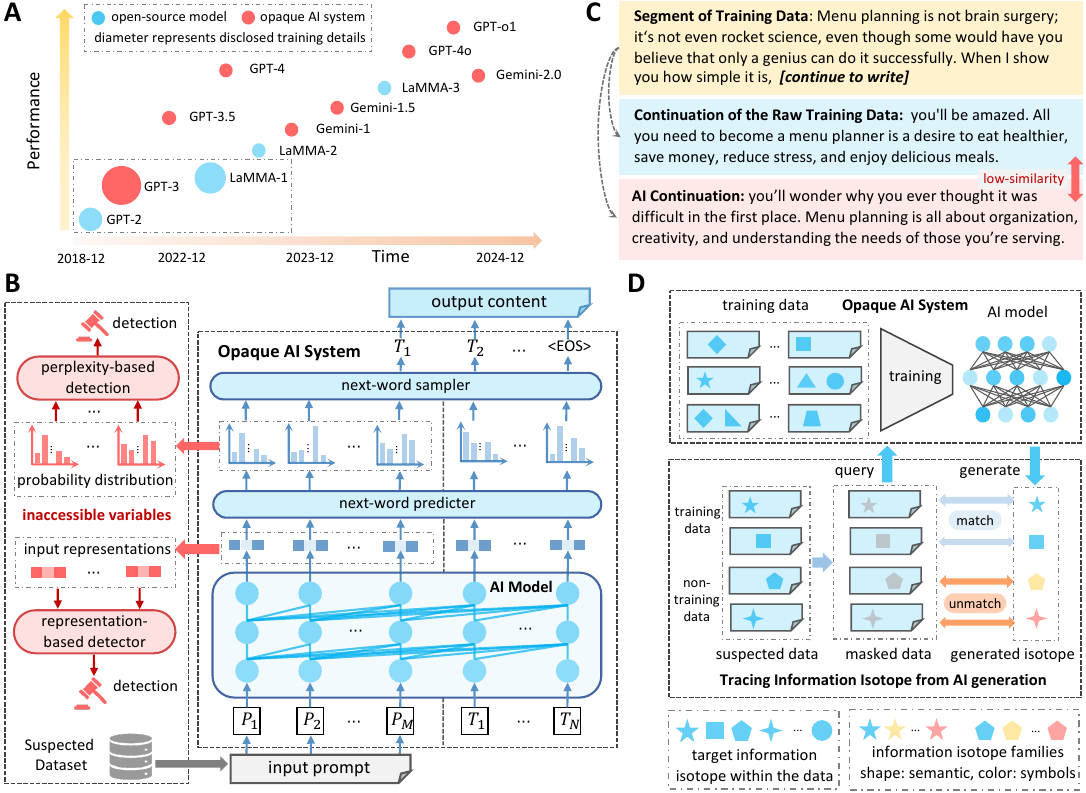}
    \caption{
The unauthorized data usage issue in AI training and corresponding detection methodologies.
\textbf{A} SOTA AI systems and their disclosure of training data details. The diameter of the circles represents the extent of disclosed information regarding training data sources. Results indicate that recent AI systems (those outside the gray box) often achieve superior performance, however, most refrain from disclosing the specific sources of their training data. This lack of transparency complicates efforts to verify unauthorized data usage in AI training.
\textbf{B} The workflow of existing methods for detecting AI training data. 
\textbf{C} A case study on detecting training data through AI-generated content. When presented with a segment of training data, the continuation generated by the AI often exhibits low similarity with the original human-authored continuation due to the advanced AI optimization algorithms. This shows the difficulty in providing conclusive evidence of training data usage through AI generations.
\textbf{D} An illustration of the traceability property of information isotopes within opaque AI systems. 
    }
    \label{fig.intro}
\end{figure}

The success of scaling laws has intensified competition among AI developers to collect ever-larger human-produced data for training purposes~\cite{longpre2024large,widder2024open}. 
However, many large-scale and high-quality datasets sourced from public internet platforms are governed by strict licensing agreements that prohibit commercial use or even data collection~\cite{longpre2024large,samuelson2023generative,heidt2024intellectual}.
Besides, the online user data collected from social media or private domains (e.g., the user interaction records on AI platforms) may also contain privacy-sensitive information, the utilization of which for AI training shall be authorized~\cite{heidt2024intellectual,beigi2020survey,ziller2024reconciling}.
Regrettably, some AI developers, either inadvertently or deliberately, incorporate unauthorized data into their training processes, thereby infringing upon the rights of data owners and raising significant ethical and legal concerns~\cite{longpre2024large,samuelson2023generative,heidt2024intellectual}. 
Moreover, AIs trained on such unauthorized data are also at an elevated risk of producing outputs that are relevant to privacy- or intellectual property-sensitive information embedded in their training data~\cite{kim2024propile,carlini2021extracting}. 
These risks not only facilitate the illegal distribution of unauthorized data, exacerbating infringements on data ownership rights, but also expose AI users to potential legal and ethical liabilities associated with data misuse.
A growing number of lawsuits, including The N.Y. Times v. Microsoft~\cite{nytms}, Tremblay v. OpenAI~\cite{treoai}, Andersen v. Stability AI~\cite{stableai}, Authors Guild v. OpenAI~\cite{authoroai}, and other similar cases, highlight the urgent need of data rights protection in AI development.

To safeguard the rights of data owners, governance institutions worldwide have enacted legislation to prevent the unauthorized use of data~\cite{voigt2017eu,goldman2020introduction,madiega2021artificial,helberger2023chatgpt}. 
A critical aspect of protecting data rights, particularly in legal disputes, lies in presenting robust evidence on data misuse.
In traditional data breach scenarios, infringers typically exploit unauthorized transactions of raw data for personal or commercial gain~\cite{alzahrani2011understanding}. 
Thus, examining the consistency between the original data and the unauthorized data can serve as compelling evidence of infringement~\cite{alzahrani2011understanding,foltynek2019academic}. 
For example, the verbatim replication of original literary work on unauthorized platforms is a clear indication of data infringement.
However, in the context of AI training, unauthorized data usage manifests differently, instead of distributing raw data, infringers mainly exploit it to enhance AI system performance~\cite{min2025watermark,zhu2024watermark,wu2024cgi,zhao2024can}.
Consequently, only the resulting AI services, rather than the original data, are made publicly accessible, posing significant challenges in identifying and evidencing data infringement~\cite{heidt2024intellectual,touvron202322llama}.
This issue is further exacerbated by the tendency of AI developers to obscure or entirely withhold information regarding the sources of their training data (Fig.~\ref{fig.intro} A).
This lack of transparency significantly hinders the ability of data owners to even become aware of unauthorized data usage in AI training, thereby complicating the enforcement of data protection regulations and the achievement of compliance with legal frameworks.

To address this challenge, several studies have investigated methods for detecting training data embedded within AI models~\cite{yeom2018privacy,shidetecting,mattern2023membership,zhangkkk,zhang2024pretraining}. 
These approaches primarily analyze the intermediate variables involved in the computational processes of target AI systems when processing the input suspected data~\cite{yeom2018privacy,shidetecting,mattern2023membership,zhangkkk,zhang2024pretraining} (Fig.~\ref{fig.intro} B).
For example, Shi et al.~\cite{shidetecting} proposed leveraging the intermediate variables reflecting model perplexity on input data, which are produced by the target AI during inference, and identifying those with low perplexities as potential instances of training data.
However, the majority of commercial AI applications are highly opaque~\cite{widder2024open}, only the AI-generated content is publicly accessible, while most internal computational variables remain inaccessible, rendering these methods impractical for many real-world implementations.\footnote{Some of these studies also discuss this limitation of their approaches when applied to mainstream opaque AI systems (such as GPT-4o and Claude) which do not provide transparent access to internal computational variables (e.g., generation logits on the input data).}
In practice, methodologies for detecting the unauthorized use of training data should be constrained to analyzing only the outputs generated by AI models. 
However, addressing this problem within such a practical framework presents significant challenges. 
State-of-the-art (SOTA) AI models possess extensive general knowledge and are explicitly designed to avoid the verbatim reproduction of their training data~\cite{chung2023increasing,lee2022factuality,huang2023large}. 
Therefore, these models may not only refuse to generate outputs that directly replicate their training data but may also produce content similar to non-training data based on their broad generalization capabilities. 
This overlap makes it inherently difficult to distinguish between outputs derived from training data and those generated from non-training data solely through output analysis.
For instance, when an AI model is prompted with a segment of its original training data, the generated continuation may exhibit low similarity to the original continuation, underscoring the complexities associated with detecting training data based on generated content (Fig.~\ref{fig.intro} C). 
Identifying training data from AI outputs needs further study.

\begin{figure}[t]
    \centering
    \includegraphics[width=0.99\textwidth]{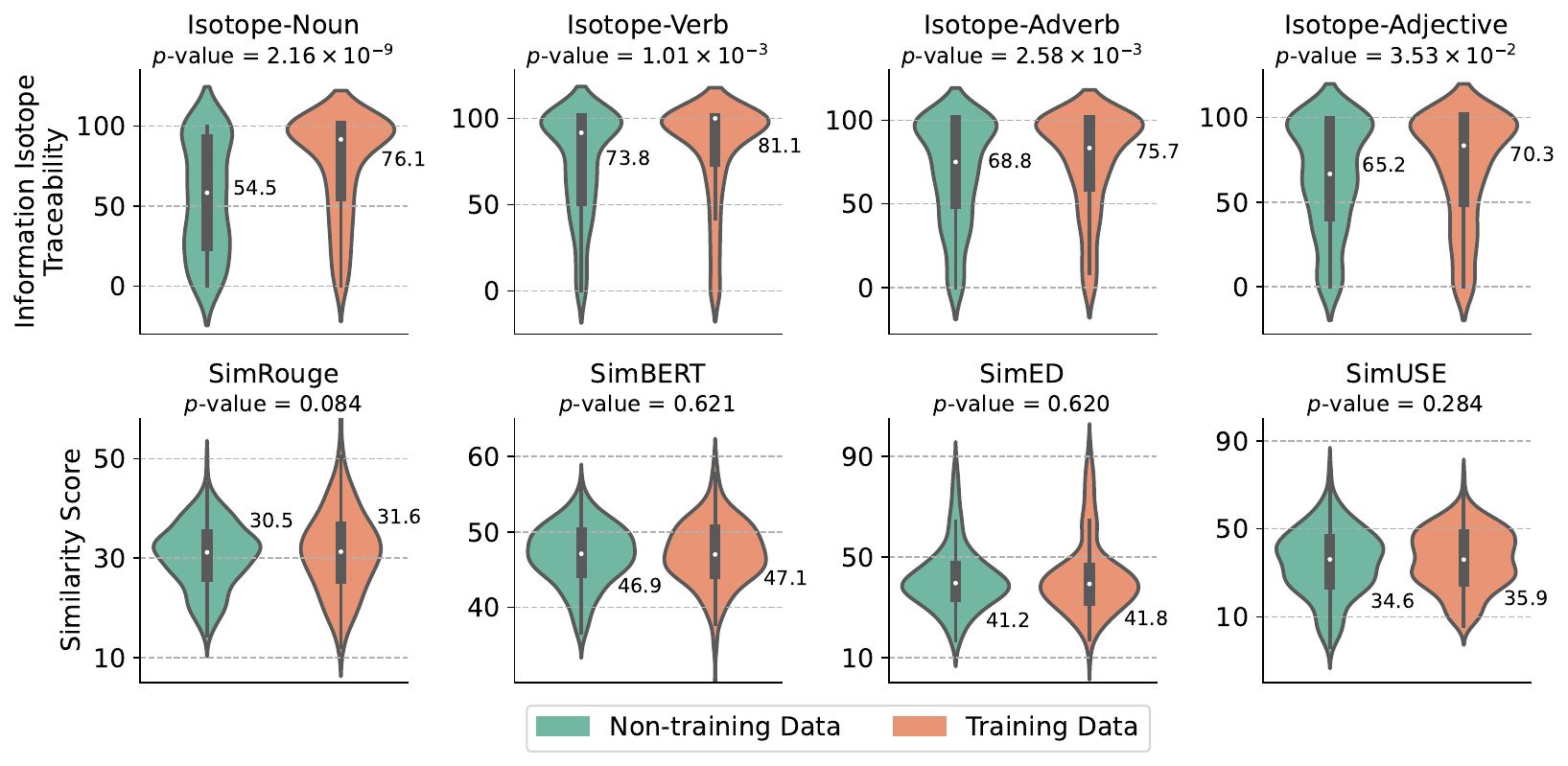}
    \caption{
    Properties of information isotopes.
    \textbf{A} Traceability distribution of information isotopes.
    This evaluation is based on the averaged results by querying six opaque AI models (including GPT-3.5, GPT-4o, Claude-3.5, Gemini-1.5, GLM-4-Air, and DeepSeek-V2.5) with copyrighted news articles. 
    The results show that information isotopes within the training dataset can be effectively traced within AI-generated content, exhibiting a traceability probability significantly higher than that observed in non-training data.
    This finding suggests that information isotopes can provide rich informativeness for distinguishing between training and non-training data based on AI-generated content.
    \textbf{B} Similarity distribution between AI-generated continuation and original training data.
    This analysis examines the similarity between the continuations generated by AI and the original training data content. 
    These results reveal that the similarity distribution between the continuations from training data and non-training data is statistically insignificant. 
    This observation highlights the inherent difficulties in identifying training data based solely on the similarity of AI-generated continuations to original training data.
    }
    \label{fig.property}
\end{figure}

In this study, we introduce the concept of information isotopes and demonstrate their traceability properties within opaque AI training systems, highlighting their potential to address this challenge.
Analogous to chemical isotopes, we identify that certain textual elements, despite their differing symbolic expression formats, consistently convey identical semantic content and naturally form a group that we define as information isotopes.
Our findings (Fig.~\ref{fig.property}) further indicate that, although these information isotopes exhibit equivalent semantic meaning in the data, a specific target expression within the training dataset is more likely to be memorized and subsequently generated by the AI when the corresponding semantic context is required.
Thus, inspired by isotope labeling techniques commonly employed to trace chemical elements in hard-to-observe microscopic scientific experiments, we propose an information isotope tracing mechanism, which can select certain information isotopes to mark training data and track their presence from AI-generated content (Fig.~\ref{fig.intro} D).
We conducted extensive experiments on six commercial LLM API services (i.e., GPT-3.5, GPT-4o, Claude, Gemini, DeepSeek, and GLM) and four open-source LLMs, using four benchmark datasets spanning three critical domains (i.e., medical texts, copyrighted books, and copyrighted news).
The results demonstrate that our method achieves remarkable detection accuracy exceeding 99\% with significant evidence ($p$-value $<0.001$), in distinguishing training data from non-training data, while baseline methods degrade to random guessing. 
Besides, our method successfully recovers information isotopes from AI-generated content with a successful rate of 75.8\%, underscoring the traceability of information isotopes in AI training. 
Further analysis highlights the robustness of our method against potential replacement-based data attack strategies and varying AI and data scales, demonstrating its generalizability across diverse real-world scenarios.
We hope this work will serve as a valuable and a general tool to assist individuals in safeguarding their data rights in both privacy and intellectual property to create an equitable environment for AI development.

\externaldocument{table/main_table}

\section*{Results}
\label{sec:Results}

\subsection*{The Property of Information Isotope}
\label{Sec.proerty}

We explore the traceability properties of the information isotope in AI training processes, to demonstrate its sufficiency and necessity as evidence for identifying unauthorized training data utilized in opaque AI systems. 
We conduct experiments using six opaque AI APIs, including GPT-4o, Claude, GPT-3.5, Gemini, DeepSeek\footnote{We access DeepSeek AI through its official API without utilizing its open-source model parameters.}, and GLM, and a dataset comprising 600 public news articles sourced from three renowned press outlets.
Half of these articles were published in 2022 (a time period commonly associated with AI training datasets) and can be found in the common crawl dataset (a public dataset widely used for training the large language models)~\cite{ccdata}, serving as the training data.
The remaining half were published in 2024, which is after the release dates of the respective AI API versions, serving as non-training data. 
We prompt the AI with the data contexts and query them to recover specific target data fragments from some ambiguous expressions sampled from the information isotope groups.
Then we calculate the recovery success rates (RSR) for training and non-training data to evaluate the traceability of information isotopes.
Fig.~\ref{fig.property} presents the results across four categories of data fragments, classified by their parts of speech.

The results demonstrate that foundation AIs effectively recover target data fragments from their information isotopes, even when these isotopes encode consistent semantics.
For instance, the average RSR across four categories of data fragments within the training dataset is approximately 75.8\%, significantly surpassing the baseline random guess rate. 
This finding underscores the memory capacity of large-scale AI models which enables them to trace specific information isotopes present in training data. 
Moreover, we observe a marked degradation in the average RSRs for non-training data compared to training data. 
For example, the averaged RSR for noun-type (NN) data fragments in the training dataset is 76.1\%, whereas for non-training data it drops to 54.5\%. 
This discrepancy is attributed to the semantic similarity among a group of information isotopes, making it significantly challenging for AIs to recover specific non-training data fragments based solely on general knowledge without relevant memory traces on the corresponding content.
These results demonstrate that information isotopes possess significant traceability properties in AI training analogous to those of isotopes used in scientific experiments. 
Further, these findings also reveal the potential of information isotopes to differentiate between training and non-training data solely through access to AI-generated content within opaque AI systems.

Next, we demonstrate that the traceability of information isotopes is not a trivial characteristic. 
To substantiate this claim, we explore an intuitive detection framework by simply analyzing the AI continuation output.
Specifically, we present the AI models with a segment of pre-existing training data prefixes and tasked them with generating the remaining content, subsequently calculating the similarity between the AI generated content and the original content of training data.
The similarity scores are measured using four metrics, including Rouge~\cite{rouge}, BERT similarity (BERT)~\cite{kenton2019bert}, edit distance (ED)~\cite{ristad1998learning}, and universal sentence  encoder (USE)~\cite{cer2018universal}.
The results in Fig.~\ref{fig.property} indicate that the average continuation similarity scores derived from training data closely approximate those from non-training data, and the corresponding statistical test suggests negligible differences between the two distributions ($p$-value $>$0.05).
This is because SOTA AIs are designed to avoid the verbatim reproduction of training data, which substantially reduces their similarities.
It proves challenging to locate AI-generated content that includes multiple consecutive words identical to those in the raw content of training data which surpass the threshold that might be considered plagiarism. 
These findings underscore that information isotopes are essential in identifying training data of opaque AI systems.
Besides, we present more analysis on the property of information isotope in Supplementary Fig. 2.

\begin{figure}[t]
    \centering
    \includegraphics[width=0.99\textwidth]{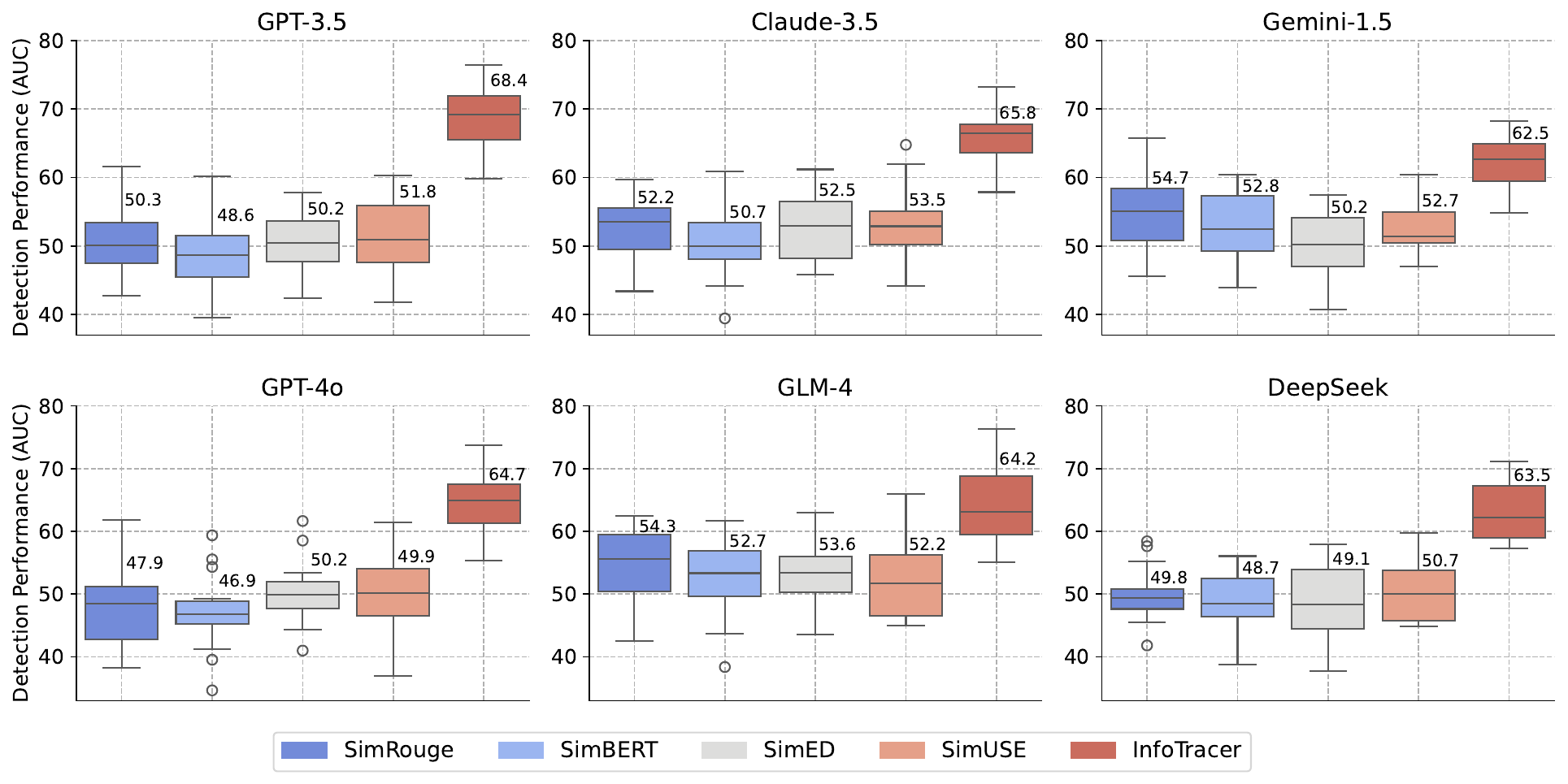}
    \caption{
    Detection performance on individual data entries.
    The evaluation was based on identifying the use of paragraphs (no more than 256 words) from news articles within six advanced AI systems. The results indicate that while baseline methods deteriorated to the level of random guessing, our method was able to effectively identify the training data, consistently and significantly outperforming the baseline approaches.
    These findings demonstrate the efficacy and superiority of our proposed information isotope tracing method in identifying training data within opaque AI systems.
    }
    \label{fig.single_detect}
\end{figure}

\begin{figure}[t]
    \centering
    \includegraphics[width=0.99\textwidth]{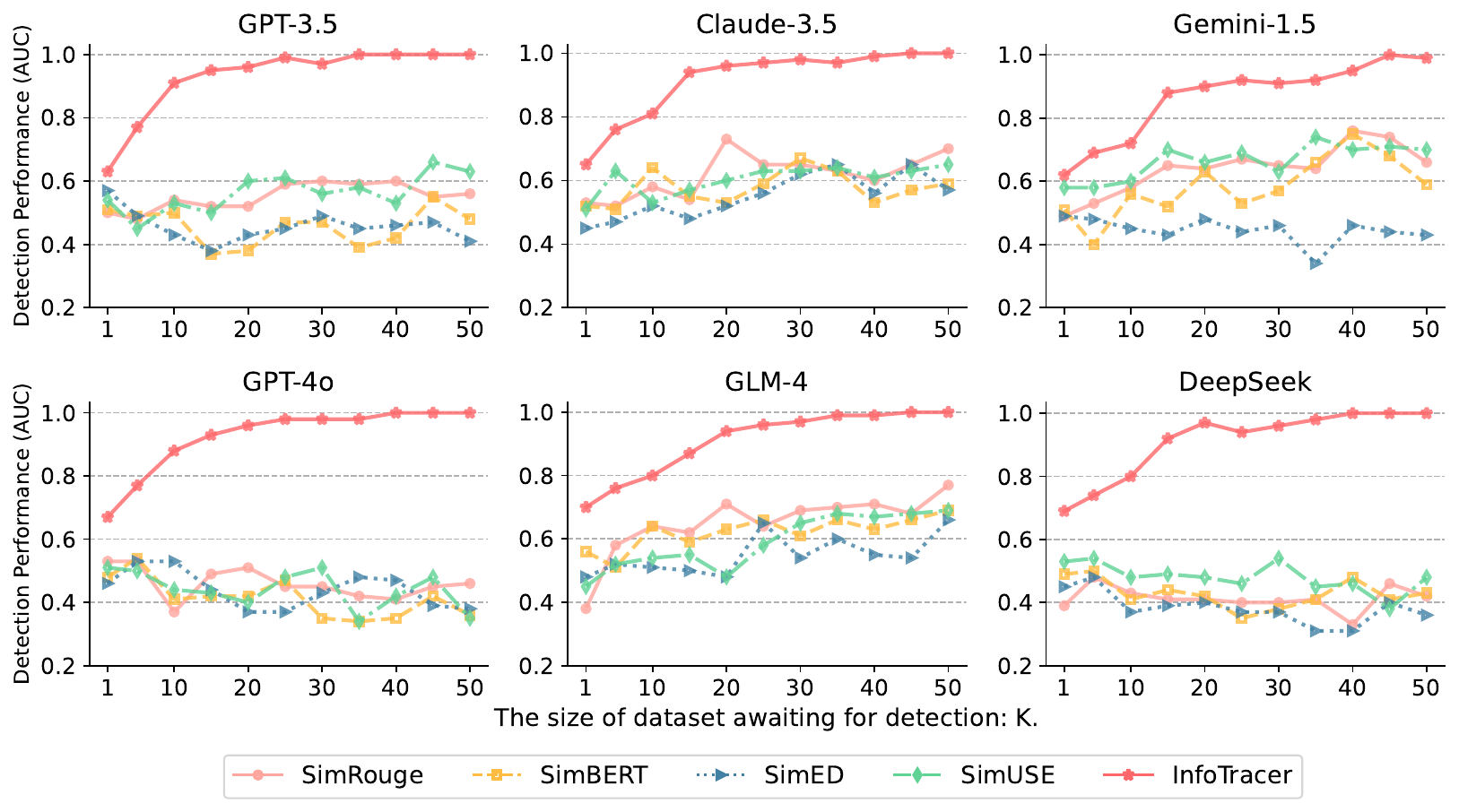}
    \caption{Detection performance under varying suspected data sizes.
    Results indicate that baselines remain ineffective in identifying training data, even when more data entries are examined.
    Instead, the performance of our method improves with increasing sizes of examined samples, and the detection accuracy of our method exceeds 99\% when only 40 data are provided. }
    \label{fig.dataset_detect}
\end{figure}

\subsection*{Detecting Training Data of Opaque AI Systems}

Next, we evaluate the effectiveness of different methods in detecting training data of opaque AIs.
Experiments are based on six widely recognized commercial super AIs and the news dataset mentioned in the previous section. 
Besides, since existing detection methods necessitate the intermediate computation variables of AI (like model perplexity on the input data), they are typically impractical in these opaque AI systems, making the corresponding evaluation also infeasible.
Thus, we adopt the continuation similarity-based methods described in previous section as the baselines of our method.
Performance on detecting the usage of a single data entry for AI training are presented in Fig.~\ref{fig.single_detect}, from which we can obtain two main findings.

First, results reveal that the baseline method performs only marginally better than random guessing. 
For instance, the accuracy of SimBERT in identifying training data from the outputs of GPT-4o and Claude is 46.9\% and 50.7\%, respectively.
This is because these advanced AIs are optimized to generate diverse content during training and minimize the likelihood of producing verbatim replications of the training data.
Besides, these super AIs are trained on vast datasets and are equipped with extensive general knowledge, enabling them to produce reasonable and contextually relevant outputs for data that was not part of their training. 
Therefore, these findings suggest that differentiating between training and non-training data based solely on the similarity between AI-generated content and original content of training data is unreliable, underscoring the challenge of conclusively identifying training data solely from AI outputs.
Second, the results demonstrate that the proposed InfoTracer method substantially enhances the performance of baseline approaches, offering significant improvements in the detection of data used in AI training. 
For example, InfoTracer achieves a detection accuracy of approximately 68.4\% against the GPT-3.5 model, representing a notable 16.6\% increase in accuracy over baseline methods.
This enhancement can be ascribed to the traceability attribute of information isotopes during AI training. 
When provided with a target content, AI models are more likely to generate data fragments that exist in their training datasets, as opposed to creating variations within the same families of information isotopes. 
The target information isotopes from training data demonstrate a significantly higher success rate in being detected within AI-generated content compared to non-training data (Fig.~\ref{fig.property}), markedly improving the informativeness for detection purposes.
Our InfoTracer method integrates an effective approach for tracing information isotopes and exhibits robust performance in identifying training data from opaque AI systems.

Next, in real-world applications, it is often more practical to identify the usage of a batch of unauthorized data rather than a single data entry in AI training. 
Thus, we further evaluate the performance of different methods in detecting the utilization of a dataset of size $K$ for AI training.
Specifically, we compute the average information isotope activities or similarity scores across entries in the target dataset as the detection metric for InfoTracer and the baseline methods, respectively.
The results, presented in Fig.~\ref{fig.dataset_detect}, yield three key observations.
First, the baseline methods still demonstrate poor efficacy in detecting the datasets used for AI training. 
For instance, when applied to the GPT-3.5 and GPT-4o models, the detection performance of the baseline methods still approximates random guessing, achieving an accuracy of around 50\%, under different data volumes. 
This finding further highlights that these baselines are inefficient in finding the evidence of data usage from AI generation, indicating they are impractical in opaque AI systems.
Second, the performance of the InfoTracer method exhibits a consistent improvement with an increasing training dataset size. 
This enhancement arises because dataset with more training data can amplify information isotope activity, thereby providing more robust signals for detection.
Third, the InfoTracer method achieves a detection accuracy exceeding 99\% across different commercial AI models when the size of the suspected dataset surpasses $40$ entries (totaling 10,000 words, equivalent to the average length of an academic paper). 
This result underscores the high sensitivity of InfoTracer, which can reliably evidence the use of training data with significant accuracy even with relatively small datasets provided. 
These findings also demonstrate the practicality and potential of InfoTracer in addressing challenging real-world tasks on the detection of small-scale unauthorized AI training data, suggesting that its generalization across real-world tasks.
In addition, we present a comprehensive analysis of the methodological design underlying our InfoTracer method, including an ablation study that evaluates the significance of two key mechanisms within InfoTracer (Supplementary Fig. 3 and 5) and a detailed examination of two important hyper-parameters (Supplementary Fig. 4 and 6). 
The results findings offer a deeper understanding on the methodology of InfoTracer. 
More details are in the Supplementary Information.

\begin{figure}[t]
    \centering
    \includegraphics[width=0.99\textwidth]{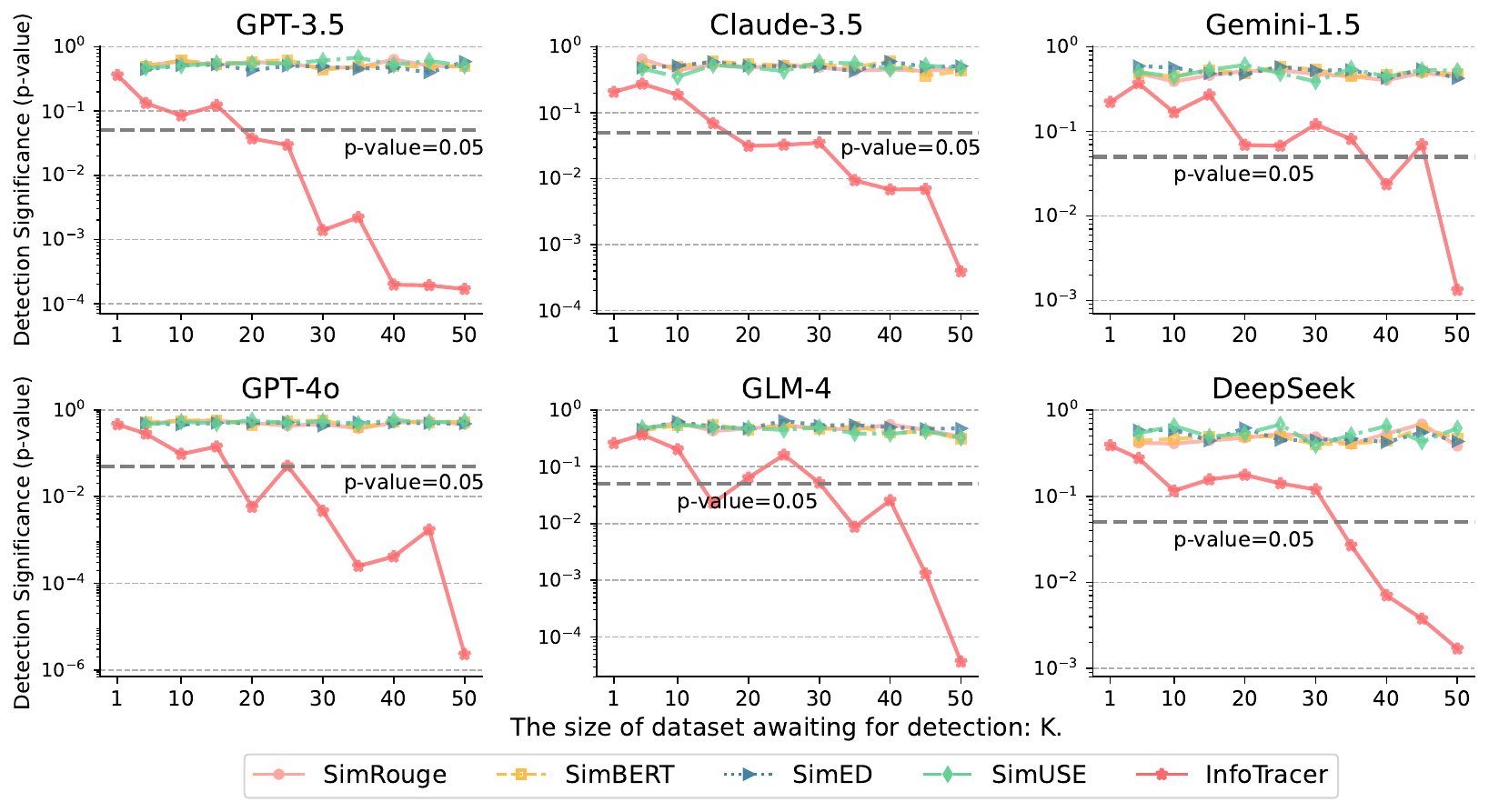}
    \caption{The statistical significance of training data detection. Results indicate that baseline detection methods exhibit negligible significance when 50 data entries are analyzed. In contrast, the detection significance of our method escalates promptly as the quantity of available data increases, illustrating its capacity to provide significant and robust evidence supporting the identifying of unauthorized data usage in AI training.
    }
    \label{fig.pvalue}
\end{figure}

\subsection*{Detection Evidence Modeling and its Evaluation}

In real-world scenarios, providing significant and robust evidence to substantiate claims of unauthorized data usage in AI training is of critical importance. 
To this end, beyond presenting the detection methodology, we also theoretically formulate the statistical significance of the detection results provided by InfoTracer (details are in the Methods section). 
Fig.~\ref{fig.pvalue} presents the statistical significance (expressed as the $p$-value) of InfoTracer compared to that of baseline methods measured by two-side t-test, under varying token volumes of the suspected data.
First, baselines show insignificant detection capability.
For instance, even when $50$ data entries with $12$k words are available for detection, the $p$-values obtained using baselines remain consistently above 0.1, failing to meet the threshold for statistical significance (i.e., \(p < 0.05\)). 
The results suggest that baselines cannot capture enough clues from the AI-generated content to distinguish between training and non-training data.
This limitation proves these baseline methods inadequate for real-world applications where robust and statistically significant evidence is essential to demonstrate unauthorized data usage in AI training.
Second, our findings reveal that InfoTracer achieves significant detection capabilities even with limited data volumes. 
Specifically, the $p$-values associated with InfoTracer consistently remain below the threshold of $0.05$ across a range of AI models when only approximately 40 data entries (10,000 words, less than the length of this paper) are available for detection.
This is because traceability properties of information isotopes from training data and non-training data are highly distinct, which can serve as the critical indicators between them.
This enables InfoTracer to extract sufficient detection clues, even for small-scale suspected data.
These results also underscore the potential of InfoTracer as a reliable tool for safeguarding small-scale individual data from unauthorized utilization by large commercial entities for AI training.
Third, as the volume of under-detected data increases, the detection significance improves at an approximately exponential rate, evidenced by the corresponding $p$-value exhibiting exponential decay. 
This phenomenon arises from the capability of our InfoTracer method to effectively amplify the informative signals embedded within the information isotopes, thereby enhancing the detection of training data.
Theoretical support for this property is provided in Lemma~\ref{lemma.1}. These findings underscore the superiority of our method in addressing critical detection scenarios involving large-scale data leakage.

\subsection*{Scalability of InfoTracer}

\begin{figure}[t]
    \centering
    \includegraphics[width=0.99\textwidth]{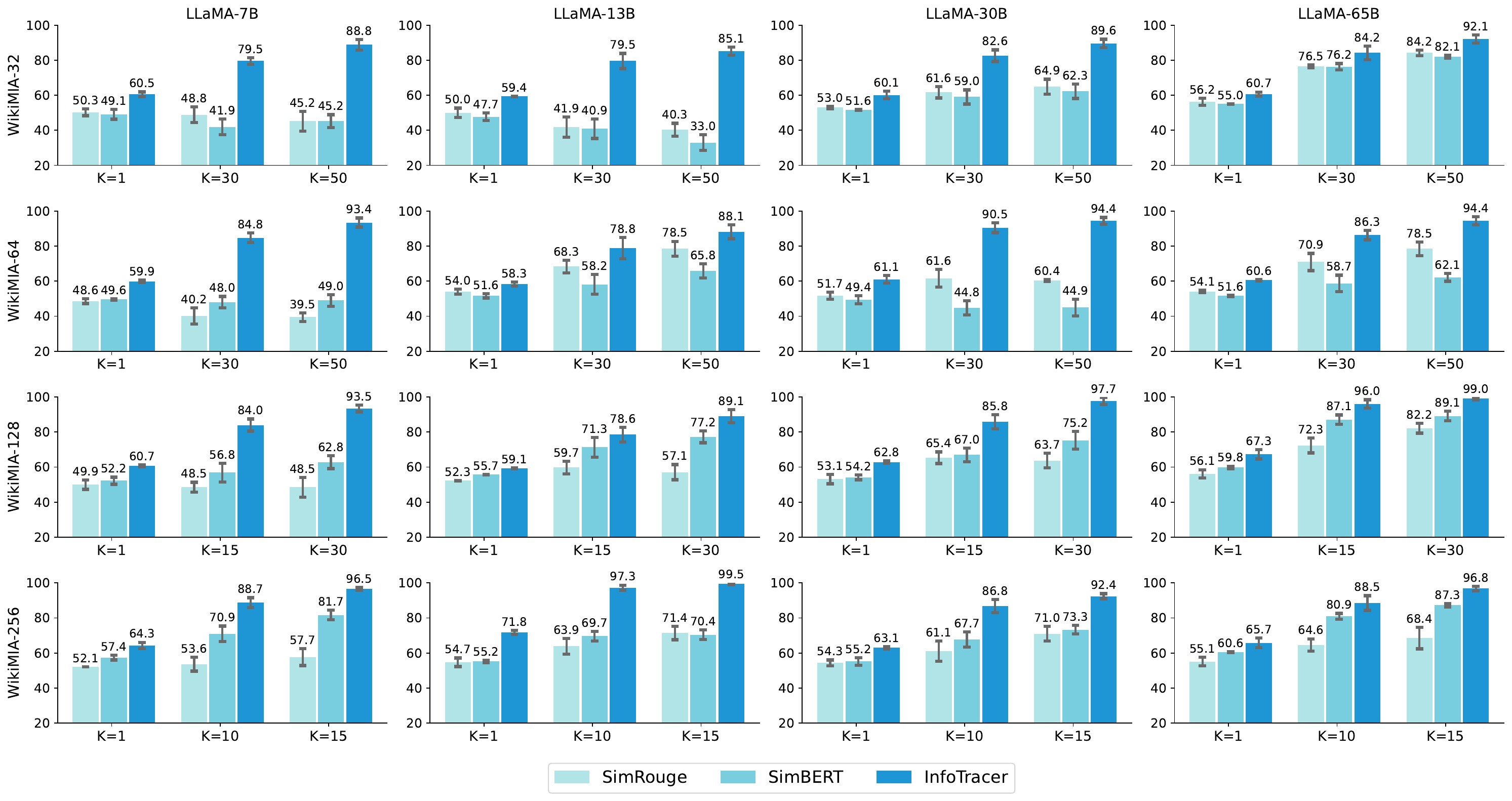}
    \caption{Scalability of InfoTracer.
    We evaluate the performance of different methods in detecting the dataset of size $K$ for AI training, under varying AI parameter scales and data lengths.
    The evaluation is based on LLaMA model and WikiMIA dataset.
    }
    \label{fig.wikimia}
\end{figure}

Next, we investigate the scalability of InfoTracer.
We evaluate its performance under varying AI model parameter sizes and detection data lengths. 
Since the exact scale of commercial AI models remains undisclosed, we utilize the open-source LLaMA series models~\cite{touvron2023llama} for evaluation.
Following the widely adopted evaluation protocol in previous works~\cite{carlini2022membership,shidetecting,zhangkkk}, the experiments are conducted on the WikiMIA dataset, a subset of which forms part of the pre-training data corpus for the LLaMA models.
The results concerning the impact of detection data lengths are summarized in Fig.~\ref{fig.wikimia}, from which two key conclusions can be drawn. 
First, we observe a significant improvement in the performance of InfoTracer as detection data lengths increase. 
This phenomenon arises because longer detection text sequences inherently contain more traceable information isotopes, which facilitate the differentiation between training and non-training data.
InfoTracer can effectively identify these useful information isotopes through the sensitive information isotope selection mechanism and leverage them to enhance the detection accuracy of training data.
Second, the results also emphasize the robust performance of InfoTracer in the challenging task of detecting short training data. 
Specifically, when identifying data entries of 32 tokens within a dataset of size $K = 50$ from the LLaMA-7B model, InfoTracer achieves an average detection accuracy of $88.8\%$, significantly outperforming baseline methods.
This can be attributed to the inherent differences in the traceability of information isotopes between training and non-training data. 
Despite the limited number of tokens in short data entries, InfoTracer can effectively capture the crtical information isotopes and trace its presence within AI generation as the detection evidence.
These results demonstrate the generalization capability of InfoTracer in detecting training data across varying input lengths, particularly when the input data length is minimal.

Besides, Fig.~\ref{fig.wikimia} also presents the performance of InfoTracer under varying AI parameter sizes.
First, we observe that as the AI scale increases, the detection performance of InfoTracer usually improves. 
This is because larger-scale AI models have greater memory capacities, which enables them to encode finer details of the training data. 
Thereby, the difference of the traceability of information isotopes against training and non-training data can be enhanced when examining larger AI models.
Besides, larger-scale AI models are typically trained on more extensive datasets, posing more risks of unauthorized data usage.
Thus, the superior performance of InfoTracer on large-scale AIs demonstrates its potential of safeguarding data rights in more critical real-world scenarios.
Moreover, the InfoTracer methodology demonstrates robust detection accuracy for training data even in smaller-scale AI models. 
For example, InfoTracer achieves a detection accuracy of 96.5\% against the LLaMA-7B model on the WikiMIA-256 dataset with $K=15$. 
This finding signifies that InfoTracer is highly sensitive in tracing the training data of AIs, thus affirming its generalizability in scenarios that deploy smaller-scale AIs, such as on-device applications.

\subsection*{Robustness of InfoTracer against Adversarial Attacks}

\begin{figure}[t]
    \centering
    \includegraphics[width=0.99\textwidth]{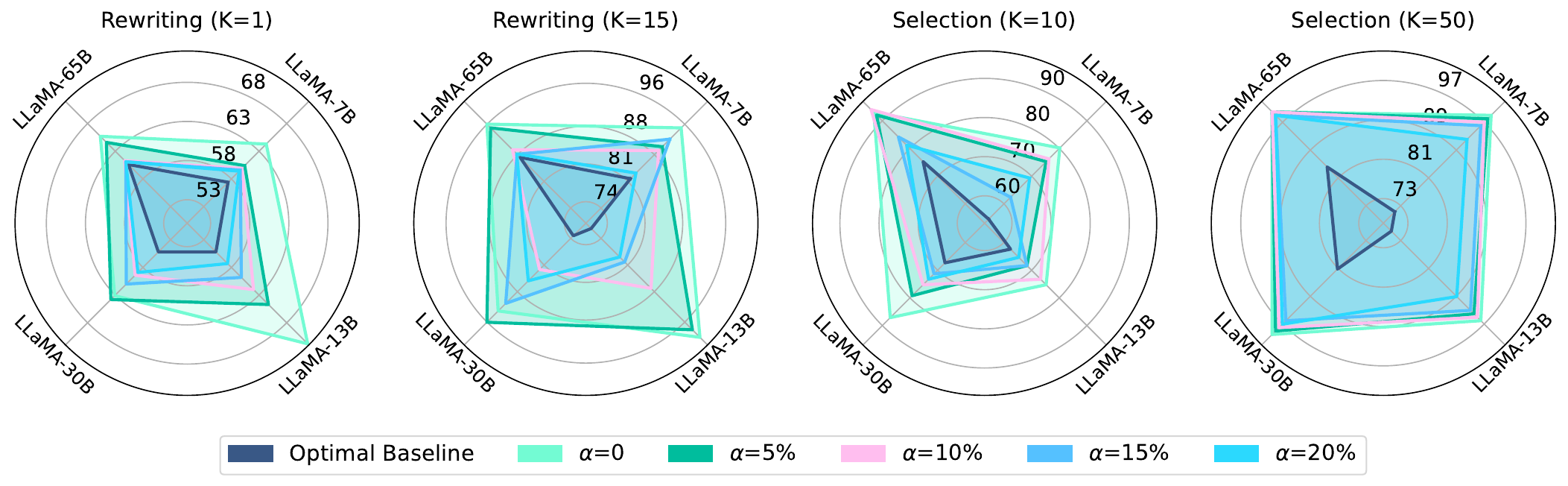}
    \caption{Robustness of InfoTracer under two replacement-based adversarial attack strategies. Rewriting-based attacks randomly replace an $\alpha$ proportion of tokens in the training data with their synonyms. 
    Selection-based attacks exclude an $\alpha$ proportion of unauthorized data from the model training process, thereby introducing noise and potentially compromising detection accuracy.
    Results based on the Wiki-128 dataset demonstrate that InfoTracer maintains its effectiveness across a range of replacement intensities, highlighting its robustness to such adversarial scenarios. }
    \label{fig.robust}
\end{figure}

\begin{figure}[t]
    \centering
    \includegraphics[width=0.99\textwidth]{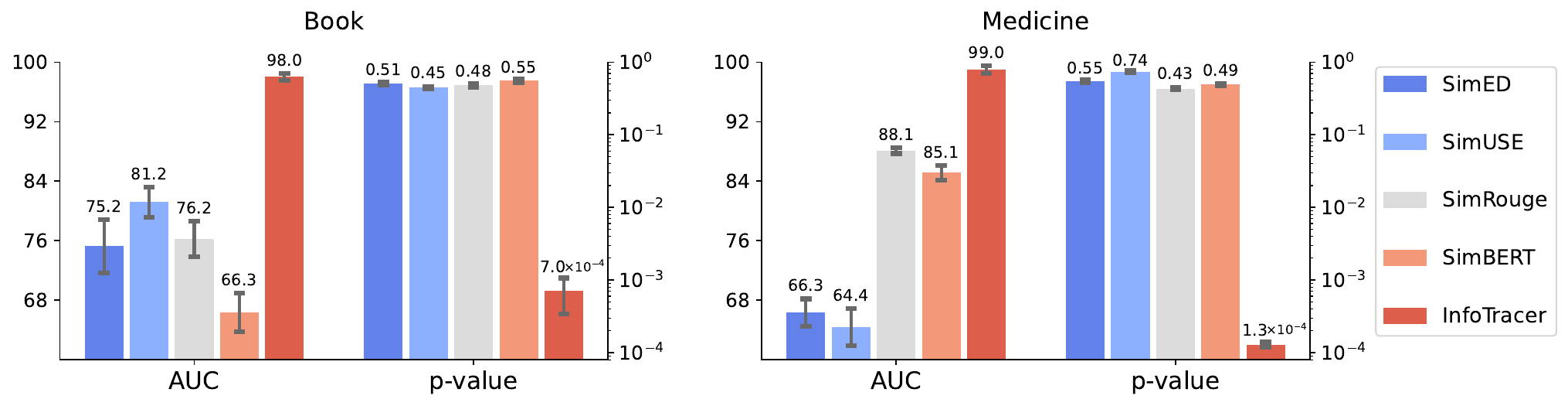}
    \caption{Generalization of InfoTracer across data distributions. We evaluate the detection performance of various methods on intellectual property-sensitive book data and privacy-sensitive medical data. The results demonstrate that InfoTracer can effectively identify training data with significant evidence, substantially and consistently outperforming baseline methods in detection accuracy and significance.
    The error bars represents the standard deviations. 
    }
    \label{fig.bookMed}
\end{figure}

The detection insight of InfoTracer lies in tracing specific information isotopes within training data, which may be susceptible to replacement-based adversarial attacks~\cite{sander2024watermarking,zhang2024remark,kirchenbauer2023watermark}.
Thus, to evaluate the robustness of InfoTracer, we systematically assess its performance under two potential replacement-based attacks: the data rewriting attack and the data selection attack.
In alignment with prior research on rewriting attack~\cite{zhang2024remark,kirchenbauer2023watermark}, we randomly sample an $\alpha$\% proportion of tokens in the training data and replace them with their synonyms. 
Besides, inspired by recent studies~\cite{sander2024watermarking}, we simulate scenarios where only a part of data from a target owner is mixed with data from other sources for AI training.
The results of these experiments, presented in Fig.~\ref{fig.robust}, reveal that replacement-based attacks of moderate intensity have a minimal impact on the performance of InfoTracer.
For instance, under the experimental setting involving the LLaMA-65B model and $K=50$ suspected data entries, the detection accuracy decreases by less than $3$\% when $20$\% of tokens are replaced.
This is because the detection evidence loss caused by the adversarial attacks can be compensated by involving some additional detection data.
Further theoretical analysis (Lemma~\ref{lemma.3}) quantifies the linear property of this trade-off, where involving an additional $2\alpha$\% of detection data entries effectively offsets the performance degradation resulting from replacing $\alpha$\% of the training data, underscoring the efficiency of InfoTracer in mitigating the effects of replacement-based attacks. 
These findings suggest that the robustness of InfoTracer can be significantly enhanced by supplementing it with a moderate volume of detection data, which is typically accessible in real-world cases.

\subsection*{Generalization of InfoTracer across Data Distributions}

Next, we evaluate the performance of InfoTracer across two critical data domains, i.e., medical data and copyrighted books, to investigate its generalizability. 
Notably, the LLaMA model and other commercial AI systems examined in prior experiments do not disclose the use of medical data in their training processes.
Thus, we employ a specialized medical large language model~\cite{cheng2023adapting} as the target AI and utilize the medical data used during its training as the target detection dataset.
Furthermore, the technical report of the LLaMA model~\cite{touvron2023llama} specifies that the Books3 dataset~\cite{rae2021scaling} is included in its training data. 
Therefore, we evaluate the performance of InfoTracer in detecting copyrighted book data using Books3 and the LLaMA model.
Results presented in Fig.~\ref{fig.bookMed} demonstrate that InfoTracer significantly and consistently outperforms baseline methods under these experimental conditions. 
For instance, InfoTracer can detect the use of medical data for AI training with an accuracy score of $99.0\%$ and a detection significance of $p=1.3\times 10^{-4}$.
These findings underscore the generalizability of InfoTracer in detecting training data across distinct critical domains of private and copyrighted data, further highlighting its effectiveness in broader real-world applications. 
Besides, we also demonstrate that InfoTracer can accurately differentiate training and non-training data with exact consistent distributions (Supplementary Fig. 1).
Detailed discussions are provided in the Supplementary Information.

Furthermore, the results presented in Fig.~\ref{fig.wikimia} also demonstrate that InfoTracer consistently and significantly outperforms baseline approaches across a wide range of experimental conditions. 
These include varying scales of target AI systems and diverse input data lengths, highlighting the generalization and superior performance of InfoTracer in satisfying diverse real-world detection needs.
In addition, the empirical evaluation presented in this section is conducted using the locally-deployed open-source LLaMA models, the local deployment of which facilitates access to computational variables during the AI inference process. 
Leveraging these variables, we evaluated the detection performance of several white-box detection methods~\cite{yeom2018privacy,carlini2021extracting}, which serve as critical benchmarks for comparison. 
We remark that the intermediate variables utilized by these methods are typically inaccessible in real-world scenarios. 
The findings in Supplementary Table 1, indicate that InfoTracer achieves performance comparable to white-box methods which can exploit inaccessible detection clues, underscoring the effectiveness of our work.

\section*{Methods}
\label{sec:Methods}

\subsection*{Problem Definition }
In this study, we investigate the method to identify the training data of a target AI model based on its generated outputs. 
Specifically, we consider an opaque AI system, denoted as $\mathcal{M}$, where the only accessible information is the model output $y$ obtained by querying it with input data $x$, i.e., $y = \mathcal{M}(x)$. 
Importantly, we assume that no additional information, including the intermediate computational variables of the AI (like model perplexity and representations of the input data), is available. 
This assumption aligns with real-world scenarios, where access to the internal workings of AI models is typically restricted.
Furthermore, given a suspected data entry $T$ of length $E$, the objective is to design a detection algorithm $\mathcal{A}(\cdot)$ capable of determining whether the data entry $T$ was used in training the AI model $\mathcal{M}$ purely through examining its generations.
In addition, in many practical scenarios, it is common for unauthorized AI training to involve a collection of data entries from a specific source rather than an isolated individual entry. 
Thus, the detection algorithm $\mathcal{A}(\cdot)$ is also tasked with analyzing a suspected dataset $\mathcal{D}$ of size $K$ to determine whether the dataset $\mathcal{D}$ was used for training the AI model, where $\mathcal{D} = \{T_i \mid i = 1, 2, \dots, K\}$.

\subsection*{Overview of the Information Isotope Tracing Method }

Inspired by the isotope labeling technique commonly employed in scientific experiments to trace microscopic matter, we propose an information isotope tracing method (termed InfoTracer) for evidencing data usage in opaque AI training processes. 
The fundamental principle of InfoTracer hinges on the traceability property of information isotopes, enabling the targeted AI model to discern and generate specific data fragments embedded within suspected training data from a set of potentially confusing information isotopes.
The core workflow of InfoTracer comprises three primary mechanisms.

First, the detection evidence contributed by an information isotope is influenced by its traceability difference across training and non-training data. 
As depicted in Fig.~\ref{fig.property}, different categories of information isotopes exhibit varying levels of traceability difference, thereby providing distinct cues for detection.
To leverage this property, we propose a sensitive information isotope selection mechanism that employs a pair of counterfactual proxy models to estimate the traceability difference of a target information isotope within a given context. 
Data fragments exhibiting significant traceability differences are selected for the subsequent information isotope labeling process, ensuring that only the most informative information isotopes are utilized for detection. 
Second, the semantic meaning of a data fragment is also shaped by its context, resulting in potential variations in its information isotopes across different contexts. 
Thus, we propose a context-aware information isotope generator, to ensure the inclusion of ambiguous information isotopes and facilitate the differentiation of the target AI in tracing training data versus non-training data.
Third, given the opacity of many cutting-edge AI systems, it is infeasible to directly access the generation probabilities of specific data fragments from their internal inference states. 
To address this limitation, we propose a selective isotope probe to estimate the recover success rate of a specific data fragment by the target AI based on its generated content.
Finally, InfoTracer models the information isotope activity for a suspected data entry by averaging the recovery success rates of the corresponding data fragments. 
This activity metric serves as a critical indicator for detecting and evidencing the usage of specific data in AI training.

\subsection*{Theoretical Analysis on the Detection Significance and Robustness of InfoTracer}

Next, we present a theoretical analysis of the detection significance and robustness of the InfoTracer method. 
Before proceeding, we reformulate the InfoTracer process from a probabilistic perspective.
InfoTracer is designed to detect whether a dataset $\mathcal{D}$, containing $K$ entries, has been used for training a target AI model.
The core process of InfoTracer involves recovering $M$ selected data fragments among its information isotopes for each data entry in $\mathcal{D}$ from the AI-generated content. 
Assume that the traceability property of information isotopes in training data and non-training data follows the distribution $\mathcal{B}(p_t)$ and $\mathcal{B}(p_n)$, respectively, where $p_t$ and $p_n$ denote the corresponding traceability probability expectation, with $p_t > p_n$. 
Then InfoTracer can observe the traceability of data fragments in dataset $\mathcal{D}$ and obtain an observation sample set $\mathcal{O}$:
\begin{equation}
    \mathcal{O} = \{ o_{i,j} \mid i = 1, 2, \dots, K, \ j = 1, 2, \dots, M \}, \quad o_{i,j} \in 
    \begin{cases}
        \mathcal{B}(p_t), & \text{if } \mathcal{Y}(\mathcal{D}) = 1, \\
        \mathcal{B}(p_n), & \text{if } \mathcal{Y}(\mathcal{D}) = 0,
    \end{cases}
\end{equation}
where $o_{i,j}$ indicates the recovery rate of the $j$-th fragment in the $i$-th data entry based on InfoTracer, and $\mathcal{Y}(\mathcal{D})$ is a binary indicator representing the usage of the dataset $\mathcal{D}$ for training the target AI.
To quantify the traceability, InfoTracer computes the information isotope activity score $\hat{p}$ by averaging the traceability outcomes of all observed  indicators: $    \hat{p} = \frac{1}{KM} \sum_{i=1}^K \sum_{j=1}^M o_{i,j}.$
% \begin{equation}
%     \hat{p} = \frac{1}{KM} \sum_{i=1}^K \sum_{j=1}^M o_{i,j}.
% \end{equation}
Finally, the computed activity score $\hat{p}$ is utilized for detecting whether the dataset $\mathcal{D}$ has been part of the AI training process.
For simplifying the subsequent analysis, we utilize $\Phi(\cdot)$ to represent the cumulative distribution function of the normal distribution, and let $N = KM$ represent the total number of selected information isotopes within the target dataset $\mathcal{D}$.

First, we give a theoretical analysis of the detection significance and performance of InfoTracer.
InfoTracer primarily identifies data usage by comparing the distribution of observed traceability outcomes, as determined by the computed activity score $\hat{p}$. 
The significance of detection in InfoTracer can be formally analyzed using the t-test, as established in Lemma~\ref{lemma.1}. 
Besides, leveraging probabilistic modeling, we can characterize the theoretical detection error of InfoTracer, as described in Lemma~\ref{lemma.2}.
These lemmas demonstrate that the detection $p$-value and the detection error associated with InfoTracer converge to zero at a super-exponential rate as the total number of utilized data fragments $N$ increases, indicating the effectiveness of InfoTracer. 
Comprehensive proofs and detailed analyses are provided in the Supplementary Information.
\begin{lemma}
\label{lemma.1}
The statistical significance $p$-value of detecting the usage of dataset $\mathcal{D}$ for AI training is given by:
\begin{equation}
    p = 1 - \Phi\left( (\hat{p} - p_n) \sqrt{\frac{N}{p_n(1-p_n)}} \right).
\end{equation}\end{lemma}

\begin{lemma}
\label{lemma.2}
The upper bound on the detection error of InfoTracer is given by:
\begin{equation}
    \text{Error} \leq \frac{1}{2(p_t - p_n) \sqrt{\pi N}} \exp\left(-(p_t - p_n)^2 N \right).
\end{equation}
\end{lemma}
Second, we evaluate the robustness of InfoTracer against replacement-based adversarial attacks designed to diminish the traceability of training data. 
Such attacks include re-writing attacks where a subset of tokens in the unauthorized data is replaced with synonyms prior to AI training, and selection attacks where only a portion of the entries from the target dataset \(\mathcal{D}\) is selectively used for AI training.
Let $\alpha$ denote the attack intensity, defined as the probability of data replacement. 
For a given detection significance level characterized by the $p$-value, the data fragment sizes required by InfoTracer under normal and adversarial scenarios are denoted by $N$ and $N'$, respectively. 
The following Lemma~\ref{lemma.3} establishes the relationship between them, which reflects the robustness of InfoTracer under adversarial attacks.
The proof of this lemma is provided in the Supplementary Information.
Based on Lemma~\ref{lemma.3}, when the attack intensity $\alpha$ is moderate (e.g., $\alpha = 0.1$), the following approximation can be derived based on the Taylor expansion: $N' \approx N (1 + 2\alpha)$.
This analysis demonstrates that the adverse effects of replacement-based attacks can be mitigated through a linear compensation in the amount of training data within $\mathcal{D}$. 
Specifically, introducing additional data entries proportional to $2\alpha N$ allows for robust detection performance.
The results indicates the robustness of InfoTracer against the potential adversarial attacks in real-world scenarios can be efficiently enhanced by incorporating a moderate of additional examined data, further demonstrating its practicality.

\begin{lemma}
\label{lemma.3}
Under the replacement-based attack scenario, InfoTracer achieves the same detection significance as in the non-attack scenario if and only if $N' = \frac{1}{(1 - \alpha)^2} N$ holds.
\end{lemma}

\subsection*{Related work}
Detecting the training data of large language models has recently emerged as an important research area~\cite{ishihara2023training, casper2024black, li2024llm, reuel2024open, carlini2021extracting}.
Most existing methods rely on the assumption that AI models exhibit distinct generation probability patterns when processing training versus non-training data~\cite{shidetecting, mattern2023membership, zhang2024pretraining}.
For instance, Yeom et al.\cite{yeom2018privacy} proposed a method that calculates the average generation probabilities of the target model for tokens in the suspected texts, and used this average as the detection score. 
Then data with higher detection scores are classified as the training data of the target AI model. 
Shi et al.\cite{shidetecting} also introduced an approach that computes the detection score based on the top K\% of tokens with the lowest generation probabilities.
This method reduces noise from high-confidence common knowledge-related tokens and amplifies the distinction between training and non-training data in detection scores.
Since AI models tend to exhibit higher confidence when generating data encountered during training, these methods often achieve satisfactory detection performance. 
However, in real-world applications, AI systems are typically highly opaque, and the generation probabilities for specific input data are often inaccessible.
Consequently, applying these methods to verify the unauthorized use of data for AI training becomes impractical.
In contrast to prior approaches, our method effectively detects training data of a target AI model with significant evidential support by solely analyzing the AI-generated outputs. It requires no access to intermediate variables, making it more applicable to real-world detection scenarios.

\section*{Acknowledgments}

This work was supported by the National Natural Science Foundation of China under Grant numbers 62425203 (S.W.), 82090053 (Y.H.), U2336208 (Y.H.).

\bibstyle{naturemag-doi}
\bibliography{cost}

\end{CJK*}
\end{document}